\documentstyle[preprint,aps,epsfig]{revtex}
\begin{document}
\tightenlines
\draft
\title{
{\begin{flushright}
{\footnotesize ZU-TH 26/01}\\
{\footnotesize YU-PP-I/E-KM-5-01}
\end{flushright}}
Decay constants, light quark masses and quark mass bounds from
light quark pseudoscalar sum rules}
\author{Kim Maltman\thanks{e-mail: kmaltman@physics.adelaide.edu.au;
permanent address: Department of Mathematics and Statistics, 
York University, 4700 Keele St., Toronto, Ontario, CANADA M3J 1P3}}
\address{CSSM, University of Adelaide, Australia 5005 \\ and}
\address{Theory Group, TRIUMF, 4004 Wesbrook Mall, Vancouver, B.C.,
CANADA, V6T 2A3}
\author{Joachim Kambor\thanks{e-mail: kambor@physik.unizh.ch}}
\address{Institut f\"ur Theor. Physik, Univ. Z\"urich,
CH-8057 Z\"urich, Switzerland}
\maketitle
\begin{abstract}
The flavor $ud$ and $us$ pseudoscalar correlators are investigated
using families of finite energy sum rules (FESR's) known to be very
accurately satisfied in the isovector vector channel.  
It is shown that the combination of constraints provided by the full
set of these sum rules is sufficiently strong to allow determination
of both the light quark mass combinations $m_u+m_d$, $m_s+m_u$ 
and the decay constants of the first excited pseudoscalar
mesons in these channels.  The resulting masses and decay constants
are also shown to produce well-satisfied Borel transformed sum rules,
thus providing non-trivial constraints on the treatment of direct
instanton effects in the FESR analysis.
The values of $m_u+m_d$ and $m_s+m_u$ obtained are in good
agreement with the values implied by recent hadronic $\tau$ decay
analyses and the ratios obtained from ChPT.
New light quark mass bounds based on FESR's involving weight functions
which strongly suppress spectral contributions from the excited
resonance region are also presented.  
\end{abstract}

\section{Introduction}
The divergence of the flavor $ij$ axial vector current in QCD is
related to the corresponding pseudoscalar density by the Ward identity
\begin{equation}
\partial_\mu A_{ij}^\mu\, =\, \left( m_i+m_j\right) \bar{q}_i i\gamma_5 q_j\ .
\label{wardps}
\end{equation}
As has been long recognized, this fact, together with the analyticity
of the correlator, $\Pi_{ij}(q^2)$, defined by
\begin{equation}
\Pi_{ij}(q^2) = i\, \int\, d^4x\, e^{iq\cdot x}\,
\langle 0\vert T\left( \partial_\mu A^\mu_{ij}(x) 
\partial_\nu {A^\nu_{ij}}^\dagger (0)\right) \vert 0\rangle
\equiv \left( m_i+m_j\right)^2\, \hat{\Pi}_{ij}(q^2) ,
\label{pscorrelator}
\end{equation}
allows one to write down sum rules 
which relate the light quark mass combinations $m_i+m_j$ 
to the decay constants of the flavor $ij$ pseudoscalar 
mesons\cite{vainshtein78}.  These sum rules, which include
the basic unsubtracted dispersion relation 
(involving $\Pi^{\prime\prime}_{ij}$, and/or its 
derivatives)\cite{vainshtein78,bnry81,gl82,npt83,d84,lrt97},
the Borel transformed version of this
relation~\cite{gl82,d84,svz,nr81,n89,bpr,jm,dps,elias97,steele00},
and finite energy sum rules~\cite{bpr,hm81,t82,kkp83,gkl84,dr87,p98},
have been used to either place bounds on $m_u+m_d$ and
$m_s+m_u$, or estimate their values.

The basic forms of these relations are, for
the unsubtracted dispersion relation (DR),
the corresponding Borel sum rule (BSR)~\cite{svz}, and finite energy sum rules 
(FESR's),
\begin{eqnarray}
&&\Pi^{\prime\prime}_{ij}(Q^2)\, =\,
2\, \int_0^\infty\, ds\, {\frac{\rho_{ij}(s)}{(s+Q^2)^3}}
\label{dispersion} \\
&&M^6\, {\cal B}\left[ \Pi^{\prime\prime}_{ij}\right](M^2)\, =\,
\int_0^\infty\, ds\, e^{-s/M^2}\rho_{ij}(s)\nonumber \\
&&\qquad\qquad\qquad\simeq\, \int_0^{s_0}\, ds\, e^{-s/M^2}\rho_{ij}(s)+
\int_{s_0}^\infty\, ds\, e^{-s/M^2}\rho^{OPE}_{ij}(s)
\label{borel} \\
&& \, {\frac{-1}{2\pi i}}
\oint_{\vert s\vert =s_0}\, ds\, w(s)\, \Pi_{ij}(s)\, = \,
\int_0^{s_0}\, ds\, w(s)\, \rho_{ij}(s)\ ,\label{basicfesr}
\end{eqnarray}
respectively, with $\rho_{ij}$ the spectral function of $\Pi_{ij}$, 
$s_0$ in line 2 of
Eq.~(\ref{borel}) the ``continuum threshold'' (beyond which
$\rho_{ij}$ is approximated by its OPE form), $M$ the Borel
mass, and $w(s)$ in Eq.~(\ref{basicfesr})
any function analytic in the region of the contour.
${\cal B}\left[ \Pi^{\prime\prime}_{ij}\right](M^2)$ in Eq.~(\ref{borel})
is the Borel transform of the OPE representation of 
$\Pi_{ij}^{\prime\prime}(Q^2)$\cite{svz}.

The LHS of either Eq.~(\ref{dispersion}) or (\ref{borel}) can
be evaluated using the OPE provided the relevant scale ($Q$ or $M$)
is large compared to the QCD scale.  For the FESR case, the 
condition that $s_0$ be similarly large is necessary, but not
sufficient, to allow reliable
evaluation of the LHS using the OPE.
The reason is that, except at extremely large $s_0$,
the OPE is expected to break down over some portion of the circle, 
$\vert s\vert = s_0$, sufficiently near the timelike real axis\cite{pqw}.
In the flavor $ud$ vector channel, where the 
spectral function has been determined very accurately from hadronic 
$\tau$ decay data\cite{ALEPH,OPAL}, one can, in fact, verify this breakdown:  
FESR's involving the weights $w(s)=s^k$ with $k=0,1,2,3$, 
which do not suppress contributions
from the region near the timelike real axis, are typically
rather poorly satisfied at scales $2\ {\rm GeV}^2<s_0<m_\tau^2$\cite{kmfesr}.
At these scales, however, this breakdown turns out to be
very closely localized to the vicinity of the
timelike axis: as soon as one restricts one's attention
to weights with even a single zero at $s=s_0$, the corresponding
FESR's are very accurately satisfied over this whole range of 
$s_0$\cite{kmfesr}.  Thus, for the ``intermediate'' scales
$2\ {\rm GeV}^2<s_0<4\ {\rm GeV}^2$ which will be of interest to us,
we must also include, as a condition for the reliability 
of the OPE representation of the LHS of Eq.~(\ref{basicfesr}),
the further requirement that $w(s_0)=0$. We will
refer to FESR's satisfying this criterion as ``pinch-weighted''
FESR's (pFESR's) in what follows.

In the region below $s\sim 4\ {\rm GeV}^2$, where the resonances
in the channels of interest ($ij=ud,us$) are well-separated,
the spectral function will be dominated by contributions
from the flavor $ij$ pseudoscalar resonances, $P$.
In the convention where $f_\pi =92.4$ MeV and
$f_K=113.0$ MeV\cite{pdg00}, the corresponding contribution to
$\rho_{ij}$, ignoring interference, is
\begin{equation}
\left[ \rho_{ij} (s)\right]_P\, =\, 2f_P^2 m_P^4 B(s)
\label{singleresonance}
\end{equation}
where $B(s)=\delta (s)$ in the narrow width approximation, with
the standard Breit-Wigner generalization to finite width,
\begin{equation}
B(s)={\frac{1}{\pi}}{\frac{\Gamma_P m_P}{[(s-m_P^2)^2+\Gamma_P^2 m_P^2]}}\ .
\end{equation}
Experimentally, both $f_\pi$ and $f_K$ are 
very accurately known, while the higher resonance ($\pi (1300)$
and $\pi (1800)$ for $ij=ud$ and $K(1460)$ and $K(1830)$ for
$ij=us$) decay constants are unknown at present.{\begin{footnote}
{$f_{\pi (1300)}$ and $f_{K(1460)}$ could, in principle, be
determined using data from hadronic $\tau$ decay, but this would
require disentangling these contributions from spin $1$ resonance
contributions in the same region.  Neither the $ud$
nor $us$ spin decomposition for the excited resonance region 
has been performed to date.}\end{footnote}}
The positivity of $\rho_{ij}(s)$, together with the fact that
the weights appearing in the spectral integrals of
Eqs.~(\ref{dispersion}) and (\ref{borel}) are $>0$, implies that
the $\pi$ (or $K$) pole contributions provide lower bounds to
these integrals.  The same is true for Eq.~(\ref{basicfesr}) as long
as the weight $w(s)$ employed is positive for $0<s<s_0$.
These lower bounds allow one to obtain corresponding 
lower bounds for $m_u+m_d$ and $m_s+m_u$\cite{vainshtein78}.
To actually determine $m_u+m_d$ and $m_s+m_u$, rather than
just set bounds on them, however, one must at present provide
theoretical input for the higher resonance contributions.
These contributions cannot be expected to be negligible since the
$f_P^2 m_P^4$ factors for all $P$ are formally of the same
order in the chiral expansion.  In fact, in existing analyses,
the higher resonance contributions are typically larger than
the $\pi$ (or $K$) pole contributions -- as an example, the $\pi (1300)$
and $\pi (1800)$ contributions to the $s^0$-weighted FESR
used to determine $m_u+m_d$ in Refs.~\cite{bpr,p98} are a factor of
$\sim 2-3$ times the $\pi$ pole contribution.

Two approaches to constraining the higher resonance contributions
exist in the literature.  In the first, additional
sum rules have been used to provide an estimate 
of the decay constant of the first excited 
resonance\cite{gl82,npt83,n89,elias97,hm81,t82,kkp83,gkl84}.  
In the second, resonance dominance has been assumed to be a good
approximation, even in the $3\pi$ (or $K\pi\pi$) threshold region, and
known ChPT expressions for the threshold values of the spectral
functions used to normalize sums-of-Breit-Wigner ans\"atze
for the higher resonance contributions.
Since the thresholds are typically several resonance
widths (or more) removed from
the resonance masses, the peak normalizations (the features of
the resonance contributions to which the sum
rule determinations of the $m_i+m_j$
are dominantly sensitive) will be ambiguous in this approach,
depending, for example, on the treatment of the 
$s$-dependence of the ``off-shell width''.
Potential dangers of this threshold normalization
approach have been discussed in Refs.~\cite{bgm98,cfnp}.
The situation in the $us$ scalar channel, where
the near-threshold behavior of the spectral function is
significantly constrained by known $K\pi$ $I=1/2$ $s-$wave
phase shifts, is particularly instructive.
As shown in Ref.~\cite{cfnp}, the near-threshold spectral
function implied, through unitarity, by the $K\pi$ phases and the 
resulting Omnes representation of the timelike scalar $K\pi$
{\it cannot} be well represented by the tail of a Breit-Wigner resonance
form; a significant background component, interfering constructively
with the resonance contribution in the threshold region,
is required.  The near-threshold
normalization of the resonance contribution is, therefore,
significantly reduced, producing a corresponding reduction in the
value of the spectral function at the $K_0^*(1430)$ resonance
peak.  This reduction is very significant numerically: the
$K_0^*(1430)$ peak value of the $us$ scalar spectral function obtained
in Ref.~\cite{cfnp} (albeit with some additional assumptions about the
high-$s$ behavior of the $K\pi$ phase and the form of the
Omnes representation) is a factor of $\sim 3$ smaller than that
obtained, using the threshold-resonance-dominance assumption (TRDA),
in Ref.~\cite{jm}.  Even if one questions the additional assumptions
which go into the precise numerical value of the reduction in this case,
one should bear in mind that the TRDA ansatz for the $us$ scalar
channel was shown to correspond to a value of the slope
of the timelike $K\pi$ form factor at threshold incompatible with
that known from ChPT~\cite{cfnp}.  Further evidence of the potential
problems of the TRDA approach are provided by the results of
Ref.~\cite{kmfesr}.  In Ref.~\cite{kmfesr}, the TRDA ans\"atze
of Refs.~\cite{bpr,p98} for the $ud$ pseudoscalar channel and 
of Refs.~\cite{jm,cps} for the $us$ scalar channel were tested 
using families of pFESR's in which the OPE scales used were the same as
those employed in the earlier analyses.
{\it If} the TRDA spectral ansatz for a given channel
is a good representation of the physical spectral function
in that channel, and if the scale of the original analysis
was such that the OPE representation could be reliably employed,
then pFESR's constructed using the same spectral ansatz for the
same correlator should also be well satisfied.  It turns out
that, in both the $ud$ pseudoscalar and $us$ scalar channels, the TRDA ansatz 
produces a very poor match between the OPE and spectral integral sides of the
various pFESR's~\cite{kmfesr}.  In contrast, the match corresponding to
the $us$ scalar spectral function of Ref.~\cite{cfnp}
is quite reasonable~\cite{kmfesr}.  

In view of the above observations, we do not employ the TRDA 
ansatz for the excited pseudoscalar contributions,
but instead constrain these contributions, in analogy to the treatment of the
isovector vector and scalar channels in Ref.~\cite{kma0}
{\begin{footnote}{Note that,
in the isovector vector channel, 
if one ignores the experimental spectral data and instead
uses the pFESR OPE integrals to fit the decay constants of a spectral ansatz
consisting of a sum of Breit-Wigner resonance contributions, one 
obtains a value of the $\rho$ decay constant in agreement with the 
experimental value to better than the experimental 
error~\cite{kma0}.}\end{footnote}}, by analyzing simultaneously
two continuous families of pFESR's, corresponding to the weights
$w^A_N(y)=(1-y)(1+Ay)$ and $w^A_D(y)=(1-y)^2(1+Ay)$, where 
$y\equiv s/s_0$.
As we will show, the set of these constraints
is sufficiently strong to allow determination of not only
the excited resonance decay constants, but also the light quark
mass combinations.  The input required for this analysis is
outlined briefly in the next section.  Our final results, together
with a discussion of existing quark mass analyses, are provided in 
Section III while Section IV contains our conclusions.

\section{Input for the pFESR analysis}

The spectral ansatz for the $ud$ pseudoscalar channel is
\begin{equation}
\rho_{ud} (s)\, =\, 2f_\pi^2m_\pi^4\delta \left( s-m_\pi^2\right)
+2f_1^2m_1^4 B_1(s) +2f_2^2m_2^4 B_2(s)\ ,
\label{rhoform}
\end{equation}
where $m_{1,2}$ are the PDG2000~\cite{pdg00} masses of 
the $\pi (1300)$ and $\pi (1800)$,
$f_{1,2}$ are their (as yet undetermined) decay constants,
and $B_{1,2}(s)$ are the standard Breit-Wigner forms.  We have employed
PDG2000 values for all resonance widths. 
The corresponding expression for $\rho_{us}(s)$ is obtained by the
replacements $\pi\rightarrow K$, $\pi (1300)\rightarrow K(1460)$
and $\pi (1800)\rightarrow K(1830)$.  In order that this ansatz
provide a good representation of the spectrum over the whole range 
required in the pFESR spectral integrals, $s_0$ cannot be taken 
much greater than $m_2^2$; if it is, an unphysical ``gap'',
with little spectral strength, will be present in the integration region.
We therefore require $s_0$ to remain less than about $(m_2+\Gamma_2)^2
\simeq 4\ {\rm GeV}^2$.  To create a good analysis window in $s_0$
without at the same time sacrificing good convergence of the
integrated $D=0$ OPE series, we also take $s_0>3\ {\rm GeV}^2$.

The ability to avoid unphysical spectral
gaps represents a potential advantage 
of the pFESR framework over its BSR counterpart.
For BSR's, the continuum threshold, $s_0$, is usually set by 
requiring an optimal stability window with respect
to the Borel mass, $M$.
Taking the $ij=us$ analysis of Ref.~\cite{dps} as an example, and considering
the case, $\Lambda_{QCD}=380\ {\rm MeV}$, which most closely corresponds
to the current experimental determination of 
$\alpha_s(m_\tau^2)$, the stability
window is optimized for $s_0$ between $6$ and $8$ GeV$^2$~\cite{dps}.  The 
resulting spectral ansatz, therefore, has a gap with very little 
spectral strength from about $5$ to $6\ {\rm or}\ 8\ {\rm GeV}^2$.
It is also worth noting that, after Borel transformation, the scale relevant
to the running coupling in the OPE is $\mu =M$.
For the correlators of interest to us the convergence of the
transformed $D=0$ series becomes good only for $M^2$ greater
than about $2\ {\rm GeV^2}$.  Even if one is willing to tolerate
a spectral gap by allowing $s_0\sim 6\ {\rm GeV}^2$, this means that
$s_0/M^2$ will be $\sim 1-2$ over much of any putative stability
window in $M$.  Such a condition signals non-trivial contributions 
from the ``continuum'' region, where only a relatively crude 
approximation to the spectral function is being employed.
This leaves only a small range of $M$ having both good OPE convergence
and acceptably small continuum contributions 
(say less than $\sim 30\%$ of the $D=0$ OPE term).
With such a small range of $M$, the BSR constraints are not
sufficiently strong to allow a simultaneous determination of the quark
masses and excited resonance decay constants.
In the case of pFESR's, empirical evidence from the isovector
vector channel suggests that contributions 
analogous to the less reliable continuum BSR contributions
({\it i.e.}, those contributions from the region 
of the contour $\vert s\vert =s_0$ near the timelike real
axis, where the OPE is expected to break down) are strongly
suppressed by the restriction to weights satisfying $w(s_0)=0$.

On the OPE side of the sum rules, one must bear in mind that,
in scalar and pseudoscalar channels, potentially important contributions 
from direct instantons exist which are not incorporated
in the standard OPE representation of $\Pi_{ij}$~\cite{novikov81}.  
Such contributions are, in fact, needed
to produce a Borel transform, ${\cal B}\left[ \hat{\Pi}_{ud}\right](M^2)$,
which behaves correctly (i.e., is independent of $M$) in the chiral 
limit\cite{novikov81,shuryak82,shuryak83,dorokhov90}.  
The instanton liquid model 
(ILM)~\cite{ilm} provides a tractable framework for estimating
such contributions.  In the ILM, an average density (related to the
value of the gluon condensate) and fixed
average size are employed for the instanton distribution.
Phenomenological constraints require the average instanton size,
$\rho_I$ to be $\simeq 1/0.6\ {\rm GeV}$\cite{shuryak82,shuryak83,ilm}.
Instanton contributions to ${\cal B}\left[ \Pi_{ud}\right](M^2)$
then exceed one-loop perturbative contributions below $M^2\sim 1\ {\rm GeV}^2$,
but drop to less than $\sim 15\%$ of this
contribution for $M^2\sim 2\ {\rm GeV}^2$~\cite{shuryak83}.
{\begin{footnote}{The combination of 2-, 3- and 4-loop contributions
roughly doubles the Borel transformed 1-loop
$D=0$ contribution at $M^2=2\ {\rm GeV}^2$, hence
further suppressing the ratio of ILM to perturbative 
contributions.}\end{footnote}}
Direct instanton contributions have been neglected in 
recent treatments of the $ud$ and $us$ pseudoscalar channels, 
apart from the BSR $ud$ analyses of Refs.~\cite{elias97,steele00}, both
of which employed the ILM.
The numerical impact of the neglect of these contributions should be
small for BSR analyses at scales $M^2>2\ {\rm GeV}^2$
since the Borel transform is known to rather strongly suppress
ILM contributions with increasing scale{\begin{footnote}{For example,
the bound obtained in Ref.~\cite{steele00} is raised by $< 5\%$ 
if ILM contributions are turned off~\cite{steeleprivate}.}\end{footnote}}.  
This is, however, {\it not} true of FESR analyses, for which ILM
contributions fall off, relative to the $D=0$ perturbative contributions,
much more slowly with increasing $s_0$.

In what follows, we will use the ILM to estimate direct 
instanton contributions to the $w^A_N$ and $w^A_D$ pFESR's.  
ILM contributions to pFESR's corresponding to polynomial weights
can be evaluated using the result~\cite{elias98}
\begin{equation}
{\frac{-1}{2\pi i}}\, \oint_{\vert s\vert =s_0} ds\, s^k
\left[ \hat{\Pi}_{ij}(s)\right]_{ILM} =
{\frac{-3\eta_{ij}}{4\pi}}\, \int_0^{s_0} ds\, s^{k+1}J_1\left(
\rho_I\sqrt{s}\right) Y_1\left(\rho_I\sqrt{s}\right)\ ,
\label{fesrinstanton}
\end{equation}
where $\eta_{ud}\equiv 1$, $\eta_{us}$ is an $SU(3)$-breaking
factor whose value in the ILM is $\sim 0.6$~\cite{shuryak83},
and the result is relevant to scales $\sim 1\ {\rm GeV}^2$.  

One should bear in mind that phenomenological support for the ILM
exists primarily for those scales ($\sim 1\ {\rm GeV}^2$) where instanton 
contributions are numerically important in pseudoscalar BSR's, 
and that this scale is significantly
lower than that ($\sim 3-4\ {\rm GeV}^2$) relevant to our pFESR analysis.
It is, therefore, useful to have an independent test of our use of the
ILM representation of instanton effects.  In this regard, one can take
advantage of the much stronger suppression of ILM contributions
in the BSR framework.  The basic idea is as follows. 
One first determines the excited meson decay constants 
for the channel of interest, using the pFESR framework.  These values
then determine the $s<s_0$ part of the
spectral ansatz for a BSR treatment of the same channel.  (The 
spectral function for $s>s_0$ is, as usual, 
approximated using the continuum ansatz; 
we fix the continuum threshold, $s_0$, following standard practice, by
optimizing the stability of the output, in this case, the quark
mass combination, $m_i+m_j$, with respect to $M^2$.)
For $M^2\sim 2\ {\rm GeV}^2$, where (1) convergence of the Borel
transformed $D=0$ series is still reasonable 
and (2) continuum contributions are still relatively small
(not yet exceeding $\sim 30\%$ of perturbative contributions), the
resulting BSR should then allow determination of
the only remaining unknown, $m_i+m_j$, with good accuracy.  
The ILM contributions play little role on the OPE+ILM side
of the BSR's at these scales, but are important for the pFESR's,
and hence for the values of the resonance decay constants
used as input to the BSR's. If the ILM representation of direct
instanton effects is reasonable at the scale of the pFESR
analysis, the pFESR and BSR determinations of $m_i+m_j$,
which will then have been 
obtained using the same excited resonance decay constants,
should be compatible within their mutual errors.  
Since the continuum approximation for the
spectral function is a relatively crude one, and the stability criterion
for choosing $s_0$ typically leaves a gap in the BSR spectral model,
there are uncertainties in the BSR analysis beyond
those associated with the uncertainties in the
OPE input, which are shared by the pFESR and BSR analyses.
In order to get a rough estimate 
of these additional uncertainties 
we allow  $s_0$ to vary in an interval of size $1\ {\rm GeV}^2$, i.e.,
by $\pm 0.5\ {\rm GeV}^2$ 
about the value corresponding to optimal stability,
and assign a $\pm 20\%$ error to the size of continuum
spectral contributions.  Since the $s_0$ values we obtain
are $>3.7\ {\rm GeV}^2$, we consider the latter estimate 
sufficiently conservative{\begin{footnote}{For the analogous cases of the $ud$
vector and axial vector channels, where the hadronic spectral
functions are known experimentally from hadronic $\tau$ decay
data, the {\it maximum} deviation of the
actual spectral function from its 4-loop OPE 
continuum approximation is less than $\sim 1/3$ of the OPE version
in the interval $2\ {\rm GeV}^2<s<m_\tau^2$~\cite{ALEPH}.
Note that these scales are smaller than those for 
which we will be employing the continuum approximation,
and that we are concerned with the {\it average}, rather than
maximum, deviation in the range $s>s_0$.}\end{footnote}}.
The uncertainties
on $m_i+m_j$ induced by use of the continuum approximation
are then not large, particularly in the
region near $M^2=2\ {\rm GeV}^2$, 
where the BSR continuum contributions are less
than $\sim 30\%$ of the $D=0$ OPE term.  The BSR/pFESR cross-check is,
as a result, most reliable at these scales.{\begin{footnote}{The ratio
of the continuum to the $D=0$ OPE contribution grows 
relatively rapidly with $M^2$.  For the $ud$ case, for
example, it has already reached $\sim 50\%$ by 
$M^2=3\ {\rm GeV}^2$ and $\sim 65\%$
by $M^2=4\ {\rm GeV}^2$.}\end{footnote}}

The OPE representation of $\Pi_{ij}^{\prime\prime}(Q^2)$ is
known up to dimension $D=6$, with the dominant $D=0$ perturbative
contribution known to 4-loop order\cite{jm,cps}.  The $D=0$ term is
given by\cite{cps}
\begin{equation}
\left[ \Pi_{ij}^{\prime\prime}(Q^2)\right]_{D=0}\, =\,
{\frac{3}{8\pi^2}}{\frac{\left( \bar{m}_i+\bar{m}_j\right)^2}{Q^2}}
\left( 1+{\frac{11}{3}}\bar{a} + 14.1793 \bar{a}^2 + 77.3683 \bar{a}^3
\right) \ ,\label{psd0}
\end{equation}
where $\bar{a}\equiv a(Q^2)=\alpha_s(Q^2)/\pi$,
$\bar{m}_k\equiv m_k(Q^2)$, with $\alpha_s(Q^2)$ and 
$m(Q^2)$ the running coupling 
and running mass at scale $\mu^2=Q^2$ in the $\overline{MS}$ scheme.
The $D=2$ term involves quark mass
corrections to the leading $D=0$ result.  For $ij=ud$
it is numerically negligible, while for $ij=us$ it is given by\cite{cps}
\begin{equation}
\left[ \Pi_{us}^{\prime\prime}(Q^2)\right]_{D=2}\, =\,
-{\frac{3}{4\pi^2}}{\frac{\left( \bar{m}_s+\bar{m}_u\right)^2
\bar{m}_s^2}{Q^4}}
\left( 1+{\frac{28}{3}}\bar{a} + \left[ {\frac{8557}{72}}
- {\frac{77}{3}}\zeta (3)\right] \bar{a}^2 \right) \ .\label{psd2}
\end{equation}
In writing Eq.~(\ref{psd2}), we have dropped terms involving
$m_{u,d}$, except in the overall prefactor $(\bar{m}_s+\bar{m}_u)^2$.
The $D=4$ $ud$ contributions are~\cite{jm}
\begin{eqnarray}
\left[ \Pi_{ud}^{\prime\prime}(Q^2)\right]_{D=4}\, &=&\,
{\frac{\left( \bar{m}_u+\bar{m}_d\right)^2}{Q^6}}
\Biggl( {\frac{1}{4}}\Omega_4
+{\frac{4}{9}}\bar{a}\Omega_3^{ss}
-\left[ 1+{\frac{26}{3}}\bar{a}\right] 2\hat{m} <\bar{u}u >
\Bigg. \nonumber \\
&&\Bigg. \qquad -{\frac{3}{28\pi^2}}\bar{m}_s^4
\Biggr)\ ,
\label{psd4ud}
\end{eqnarray}
where $\Omega_4$ and $\Omega_3^{ss}$ are the RG invariant modifications
of $\langle a G^2\rangle$ and $\langle m_s\bar{s}s\rangle$ defined
in Ref.~\cite{jm}, $\hat{m}=(m_u+m_d)/2$, and we have dropped numerically 
negligible terms of $O(\hat{m}^4)$; the $D=4$ $us$ contributions are, 
similarly,~\cite{jm}
\begin{eqnarray}
\left[ \Pi_{us}^{\prime\prime}(Q^2)\right]_{D=4}\, &=&\,
{\frac{\left( \bar{m}_s+\bar{m}_u\right)^2}{Q^6}}
\Biggl( {\frac{1}{4}}\Omega_4
+\left[ 1+{\frac{64}{9}}\bar{a}\right]\Omega_3^{ss}
-2<m_s\bar{u}u>\left[ 1+{\frac{23}{3}}\bar{a}\right]
\nonumber \Bigg. \\
\Bigg.&&\qquad
-{\frac{3}{7\pi^2}}\left[ {\frac{1}{\bar{a}}} +{\frac{155}{24}}\right] 
\bar{m}_s^4\Biggr)\ ,
\label{psd4us}
\end{eqnarray} 
where we have again
dropped terms suppressed by powers of $\hat{m}/m_s$ relative
to those shown, except in the overall $ (\bar{m}_s+\bar{m}_u)^2$
prefactor.
Finally, the $D=6$ contributions are~\cite{jm}
\begin{eqnarray}
\left[ \Pi_{ij}^{\prime\prime}(Q^2)\right]_{D=6}\, &=&\,
{\frac{\left( \bar{m}_s+\bar{m}_u\right)^2}{Q^8}}
\Biggl( -3\left[
\langle m_ig\bar{q}_j\sigma\cdot G q_j+m_jg\bar{q}_i\sigma\cdot G q_i\rangle
\right]\nonumber \Bigg. \\
\Bigg. &&\qquad -{\frac{32}{9}}\pi^2 a\rho_{VSA}
\left[ \langle \bar{q}_i q_i\rangle^2
+\langle \bar{q}_j q_j\rangle^2 - 9\langle \bar{q}_i q_i\rangle
\langle \bar{q}_j q_j\rangle\right]\Biggr)\ ,
\label{psd6}
\end{eqnarray}
where $\rho_{VSA}$ describes the deviation of the four-quark condensates
from their vacuum saturation values.

Numerical values of the input required on the OPE+ILM side of the 
sum rules are as follows:  
$\rho_I=1/(0.6\ {\rm GeV})$~\cite{shuryak83,ilm},
$\alpha_s(m_\tau^2)=0.334\pm .022$~\cite{ALEPH,OPAL},
$\langle \alpha_s G^2\rangle = (0.07 \pm 0.01)\ {\rm GeV}^4$~\cite{narisonaGG},
$\left( m_u+m_d\right)\langle \bar{u}u\rangle =-f_\pi^2 m_\pi^2$ (the
GMOR relation)
{\begin{footnote}{Deviations from the GMOR relation have recently been shown 
to be at most $6\%$ \cite{CGL01}. The resulting error on the 
$m_s$ analysis is completely negligible.}\end{footnote}}
, $0.7< \langle \bar{s}s\rangle /\langle \bar{u} u\rangle 
\equiv r_c<1$~\cite{jm,cps};
$\langle g\bar{q}\sigma Fq\rangle
=\left( 0.8\pm 0.2\ {\rm GeV}^2\right)\langle \bar{q} q\rangle$\cite{op88}
and $\rho_{VSA}=5\pm 5$ ({\it i.e.}, allowing, to be conservative,
up to an order of magnitude 
deviation from vacuum saturation for the four-quark condensates).
The $D=0,2$ and $4$ contributions to the OPE integral have been evaluated 
using contour-improvement~\cite{cipt1,cipt2}, which is known to
improve convergence and reduce residual scale dependence~\cite{cipt2}.
For this purpose, we employ the analytic solutions for the running
coupling and running mass obtained using the 
known 4-loop-truncated versions of the $\beta$~\cite{beta4} 
and $\gamma$~\cite{gamma4} functions, with the value of $\alpha_s(m_\tau^2)$
noted above as input. 

\section{Results and Discussion}
\subsection{Quark Mass Bounds}
Bounds for the light quark masses based on the known values of
the $\pi$ or $K$ pole contributions and the positivity of the
spectral function, whether obtained using 
the dispersion formulation, BSR's or FESR's, all depend on
the scale employed in the OPE.  Since, at the scales for which the
resulting bounds are of phenomenological interest, the
$O(a^2)$ and $O(a^3)$ terms in the integrated
$D=0$ OPE series are not numerically negligible, earlier versions
of these bounds, based on two-loop and three-loop forms 
of the $D=0$ part of $\Pi_{ud,us}$, are superceded by
the work of Ref.~\cite{lrt97} (LRT), which employed the 4-loop
OPE expression.  The bounds of LRT are based on the dispersion
relation for $\Pi_{ij}^{\prime\prime}$, and the higher derivative
moments thereof.  Restricting our attention to the results in LRT
corresponding most closely to the experimental value
$\alpha_s(m_\tau^2)=.334$, {\it i.e.}, $\Lambda^{(3)}_{QCD}=380\ {\rm MeV}$, 
the most stringent bounds arise from what in LRT is called
the ``quadratic inequality''~\cite{lrt97}.
These bounds decrease with increasing OPE scale, $Q^2$, and,
for $Q^2=4\ {\rm GeV}^2$, yield (from Figs. 2 and 3 of LRT)
\begin{eqnarray}
&&[m_s+m_u](\mu =2\ {\rm GeV})> 105\ {\rm MeV}\nonumber \\
&&[m_u+m_d](\mu =2\ {\rm GeV})> 8.1\ {\rm MeV}\ .
\label{psbounds}\end{eqnarray}
Normally one would expect the convergence of the 4-loop 
$D=0$ OPE series to be quite good at scales as large as 
$Q^2=4\ {\rm GeV}^2$.  In this case, however, 
the denominator appearing on the RHS of the
quadratic bound (see Eq. (19) of LRT), which has the form
\begin{equation}
\left[ 3 {\cal F}_0^{QCD} {\cal F}_2^{QCD}-
2\left( {\cal F}_1^{QCD}\right)\right] =1+{\frac{25}{3}}\bar{a}
+61.79 \bar{a}^2+517.15 \bar{a}^3+\cdots \ ,
\label{lrtdenom}\end{equation}
is very slowly converging,
behaving as $1+.83+.61+.51$ at $Q^2=4\ {\rm GeV}^2$.
The bounds in Eq.~(\ref{psbounds}) are thus likely to 
have a significant residual uncertainty associated with 
the truncation at $O(a^3)$
{\begin{footnote}{If one wished to 
work, e.g., at a scale such that the $O(a^3)$
term in Eq.~(\ref{lrtdenom}) were less than $\sim 20\%$ 
of the leading term, one
would need to go to $Q^2\sim 9\ {\rm GeV}^2$, at which scale
the bounds on $[m_s+m_u](\mu =2\ {\rm GeV})$ and
$[m_u+m_d](\mu =2\ {\rm GeV})$ would be reduced to $\sim 60$
and $\sim 3.4$ MeV, respectively.}\end{footnote}}.
The behavior of the $D=0$
series is in fact much better for the zeroth moment LRT bound.
At the lowest scale shown in Fig.~1 of LRT ($Q=1.4\ {\rm GeV}$), 
the behavior of the $D=0$ series is $1+.45+.22+.15$,
already quite well-converged.  The corresponding bound on
$m_s+m_u$ which, reading from Fig.~1, is
\begin{equation}
[m_s+m_u](\mu =2\ {\rm GeV})> 80\ {\rm MeV}\ ,
\label{lrtmsmusafebound}
\end{equation}
thus seems to us to be subject to significantly less truncation-induced
uncertainty.  Although the zeroth moment bound for
$m_u+m_d$ is not quoted in LRT, the result of Eq.~(\ref{lrtmsmusafebound}),
together with the result $R\equiv 2m_s/(m_u+m_d)=24.4\pm 1.5$ determined
from ChPT~\cite{leutwylerqmasses}, would imply 
\begin{equation}
[m_u+m_d](\mu =2\ {\rm GeV})> 6.6\ {\rm MeV}\ .
\label{lrtmdmusafebound}
\end{equation}
The result of Eq.~(\ref{lrtmdmusafebound})
is in good agreement with the bound obtained 
by the same authors~\cite{lrt97}
from the study of the $ud$ scalar channel{\begin{footnote}{The 
$D=0$ OPE series corresponding
to this bound has the same (good) convergence behavior
as that given for the zeroth moment
bound above.}\end{footnote}}
using constraints on the timelike scalar-isoscalar $\pi\pi$ form factor
from ChPT and $\pi\pi$ phase shift data in the region
$4m_\pi^2 < s < (500 \ {\rm MeV})^2$ 
(see Fig.~4 of LRT),
\begin{equation}
[m_u+m_d](\mu =2\ {\rm GeV})> 6.8\ {\rm MeV}\ .
\label{lrtmdmusafebound2}
\end{equation}
An analogous bound for $m_s$ was obtained from a treatment of the 
$us$ scalar correlator employing ChPT constraints for the timelike
scalar $K\pi$ form factor~\cite{ls98}.  Taking the case 
from that reference corresponding
to the plausible assumption that the one-loop ChPT expression for
the $K\pi$ form factor is accurate to the $.5-1\%$ level in the region
$0<s<m_K^2-m_\pi^2$, the resulting bound is
\begin{equation}
m_s(\mu = 2\ {\rm GeV})>65\ {\rm MeV}\ ,
\label{ls98bound}
\end{equation}
which is less stringent than that in Eq.~(\ref{lrtmsmusafebound}).
Other recent bounds are 
(1) that obtained in Ref.~\cite{dn98} by combining the
upper bound on $\langle \bar{q}q\rangle (1\ {\rm GeV})$ 
allowed by the analysis of the 
$D\rightarrow K^*\ell\nu_\ell$ vector form factor
with the (assumed to be well-satisfied) GMOR relation,
\begin{equation}
[m_u+m_d](2\ {\rm GeV})>6.8\ {\rm MeV}\ ,
\label{dn98mumdbound}\end{equation}
and (2) that obtained in Ref.~\cite{steele00}
using BSR's and H\"older inequalities at scales $\sim m_\tau$
\begin{equation}
[m_u+m_d](2\ {\rm GeV})>4.2\ {\rm MeV}\ .
\end{equation}
Note that the latter bound was obtained including ILM contributions
on the OPE side of the sum rule; the bound is $\sim 5\%$ 
higher if ILM contributions are turned off~\cite{steeleprivate}.
All other bounds noted above were obtained neglecting direct
instanton contributions.  This neglect should have little
impact on dispersive bounds such as that of Eq.~(\ref{lrtmsmusafebound})
since, if one uses the ILM to estimate these effects,
the lower bound of Eq.~(\ref{lrtmsmusafebound}) is reduced
by only $3\ {\rm MeV}$.

An alternate approach to using the positivity of $\rho_{ij}$
to obtain quark mass bounds is to employ pFESR's with weights
satisfying $w(s)\geq 0$ in the region $0<s<s_0$.  A potential
advantage of this approach is the freedom to choose weights
which strongly suppress contributions from the excited resonance region.
Strong suppression of this type should lead to
bounds which are ``close'' to the actual mass values.
One can arrange such strong suppression by choosing
$w(y)=(1-y)^N p(y)$ with $N$ sufficiently large.  Here
$p(y)$ is a ``residual polynomial'' which has to be
chosen in such a way as to (1) keep the coefficients in $w(y)$ small 
(thus avoiding the growth of unknown higher $D$ contributions)
{\begin{footnote}{Without this constraint, working with high
powers of the factor $(1-y)$ typically produces 
polynomials with large coefficients for the higher degree $y^k$
terms. Since $y^k$ terms with large $k$ are associated with
OPE contributions of large dimension, which are poorly constrained 
phenomenologically, large $y^k$ coefficients
signal potentially large, and essentially unknown,
non-perturbative contributions~\cite{ckp98,gkp01}, and
hence must be avoided.}\end{footnote}}
and (2) retain
good convergence of the integrated $D=0$ OPE series.
The construction of such weights was considered in a different
context previously~\cite{km00}.  Here we consider quark mass bounds
based on pFESR's employing the three weights of this type
constructed in Ref.~\cite{km00}.  It turns out that both the $D=0$ OPE
convergence and the stringency of the resulting bounds is 
best for the case of the weight called $w_{20}(y)$ in Ref.~\cite{km00},
so we present results only for this case.  The behavior of $w_{20}(y)$
in the integration region ($0<y<1$) is shown in Fig.~1.  (Its
explicit form may be found in Ref.~\cite{km00}.)
For $s_0=4\ {\rm GeV}^2$, the contour-improved $D=0$ OPE series
for the $w_{20}$ pFESR, truncated at $O(a^3)$, converges quite reasonably, 
behaving as $\sim 1+.55+.28+.19$.  Moreover,
since, for example, if we define 
$y_{K(1460)}\equiv m^2_{K(1460)}/4\ {\rm GeV}^2$, 
$w_{20}\left(y_{K(1460)}\right)=0.11$,
there will be nearly an order of magnitude suppression
of excited resonance contributions, relative to the $K$
contribution, in the $us$ channel.  Unfortunately, the $D=0$
convergence deteriorates if one tries to go to lower $s_0$,
where this suppression would be much stronger.
Ignoring possible direct instanton contributions, one obtains
\begin{equation}
m_s(2\ {\rm GeV})> 93\ {\rm MeV}\ .
\label{km00msbound}\end{equation}
The convergence is obviously not sufficiently rapid that one
should rule out values of the bound a further $\sim 5$ or so MeV lower.
The analogous bound for $m_u+m_d$ is
\begin{equation}
[m_u+m_d](2\ {\rm GeV})>6.6\ {\rm MeV}\ .
\label{km00mumdbound}
\end{equation}
These bounds should be compared only to those bounds listed
above which also neglect possible instanton effects.
As expected, the rather strong suppression of
excited resonance contributions relative to $K$
pole term produces a bound on $m_s$, Eq.~(\ref{km00msbound}), which is more 
stringent than the zeroth moment
LRT bound.  The $m_u+m_d$ bound of Eq.~(\ref{km00mumdbound}), however,
remains comparable to the LRT $m_u+m_d$ bound, though still having
the advantage that one would expect it to represent a better
approximation to the true value.
If one now incorporates an estimate of direct instanton effects
using the ILM, the bounds of Eqs.~(\ref{km00msbound})
and (\ref{km00mumdbound}) are reduced to
\begin{equation}
m_s(2\ {\rm GeV})> 84\ {\rm MeV}
\label{msilmbound}\end{equation}
and
\begin{equation}
[m_u+m_d](2\ {\rm GeV})>5.7\ {\rm MeV}\ .
\label{mumdilmbound}\end{equation}
The bound of Eq.~(\ref{msilmbound}) remains slightly more stringent than
that of Eq.~(\ref{lrtmsmusafebound}).  A more stringent bound on
$m_u+m_d$,
\begin{equation}
[m_u+m_d](2\ {\rm GeV})>6.9\ {\rm MeV}\ ,
\label{finalmumdbound}\end{equation}
can be obtained using Eq.~(\ref{msilmbound}) in combination
with the mass ratios obtained from ChPT~\cite{leutwylerqmasses}.

To go beyond these bounds, we must attempt to also
determine the excited resonance decay constants as part of
the pFESR analysis.  This extension of the analysis is described in
the next section.

\subsection{Quark Masses and Excited Meson Decay Constants}

To simultaneously extract $m_i+m_j$ and the corresponding
excited pseudoscalar decay constants, we perform a combined
analysis of pFESR's based on the weight families 
$w^A_N(y)$ and $w^A_D(y)$ and work in the window 
$3\ {\rm GeV}^2\leq s_0 \leq 4\ {\rm GeV}^2$.  For these $s_0$,
the $D=0$ OPE series converges well for all $A\geq 0$,
and the spectral ansatz should be of the correct
qualitative form.  Larger values of $A$ correspond 
to larger relative contributions
from the excited resonance region, and hence are useful for
constraining the unknown resonance decay constants.
To explore sensitivities to the choice of analysis regions,
we have also considered the alternate ranges
$3.6\ {\rm GeV}^2\leq s_0 \leq 4\ {\rm GeV}^2$ and $2\leq A\leq 6$, as
well as considering separate $w_N^A(y)$ and
$w_D^A(y)$ analyses (thus checking the mutual consistency
of the pFESR's corresponding to the two weight families).
The only significant impact
of uncertainties in the experimental input for the resonance
parameters is that occurring in the $ud$ analysis, 
associated with the $\pi (1300)$ width;
this is a consequence of the rather wide range, 
$200< \Gamma \left(\pi (1300)\right) < 600$ MeV given in the
PDG2000 compilation.  In what follows, we quote errors from
this source separately, labelling them with the subscript ``$\Gamma$''.
Uncertainties associated with changes in the 
$s_0$ and $A$ analysis windows and weight family choice
are added in quadrature and denoted by the subscript
``method''.  Finally, those errors denoted by the subscript ``theory'' 
are obtained by combining in quadrature errors associated
with uncertainties in the OPE input parameters
$\rho_{VSA}$, $\langle \alpha_s G^2\rangle$,
$\alpha_s(m_\tau^2)$ and $r_c$ and our estimate of the error due to
truncation of the dominant $D=0$ OPE contribution at 4-loop
order.  The latter is obtained by evaluating the $O(a^4)$
contribution that would result if we assumed continued geometric
growth of the coefficients, {\it i.e.}, the presence of an additional term
$\sim 422 \bar{a}^4$ in the polynomial factor of 
Eq.~(\ref{psd0}).{\begin{footnote}{In view of the discussion in
Section 5 of Ref.~\cite{kataev01}, this estimate is likely to
be a very conservative one.}\end{footnote}}
It turns out that, when ILM contributions are included, 
the spectral contribution of the second resonance is
small for both channels, and hence that the corresponding decay constant
can be determined with only limited accuracy.
When quoting results for the second decay constant in this case, we 
will, therefore, display only the range of values allowed by
the combined (i.e., ``theory'', ``method'' and (for the $ud$ channel only) 
``$\Gamma$'') errors. The analysis of the $ud$ channel has been described 
briefly already in Ref. \cite{MKlett01}.

The results obtained from the analysis, when ILM contributions
are included on the theoretical side of the pFESR's, are as follows.
For the $ud$ channel we have
\begin{eqnarray}
&&[m_u+m_d](2\ {\rm GeV})\, =\, 7.8\pm 0.8_\Gamma 
\pm 0.5_{theory}\pm 0.4_{method}\ 
{\rm MeV}\label{udilmmass} \\
&&f_{\pi (1300)}\, =\, 2.20\pm 0.39_{\Gamma}\pm 0.18_{theory}\pm 0.18_{method}
\ {\rm MeV}\label{udilmf1} \\
&&0< f_{\pi (1800)}< 0.37\ {\rm MeV}\ ,
\label{udilmf2}
\end{eqnarray}
and for the $us$ channel
\begin{eqnarray}
&&m_s(2\ {\rm GeV})\, =\, 100\pm 4_{theory}\pm 5_{method}\ 
{\rm MeV}\label{usilmmass} \\
&&f_{K(1460)}\, =\, 21.4\pm 1.6_{theory}\pm 2.3_{method}
\ {\rm MeV}\label{usilmf1} \\
&&0< f_{K(1830)}< 8.9\ {\rm MeV}\ .
\label{usilmf2}
\end{eqnarray}
From Eqs.~(\ref{udilmmass}) and (\ref{udilmf1}), we 
see that the uncertainty in the $\pi (1300)$ width is,
in fact, the dominant source of error
in the determination of both $m_u+m_d$ and $f_{\pi (1300)}$.
To get a feel for the
relative size of the various contributions to the ``theory''
error we note that, for the $ud$ case, the errors in
$[m_u+m_d](2\ {\rm GeV})$ due to
the uncertainties noted above on the input parameters $\rho_{VSA}$, 
$\langle \alpha_s G^2\rangle$, $\alpha_s(m_\tau^2)$ and
truncation at $O(a^3)$ are $\pm 0.25$, $\pm 0.05$,
$\pm 0.28$ and $\pm 0.25$ MeV, respectively.  The corresponding
contributions to the errors on $m_s(2\ {\rm GeV})$ are
$\pm 1.5$, $\pm 0.4$, $\pm 2.3$ and $\pm 3.1$ MeV, respectively,
with a further contributions of $\pm 0.2$ MeV due to the
range of $r_c$ employed in this case.
The agreement between the OPE and spectral integral sides of the 
various pFESR's corresponding to the results above
is very good. 
The fit quality for the $us$ channel is
displayed, for the $w^A_N$ and $w^A_D$ families, 
in Figs.~\ref{figusNILM} and \ref{figusDILM}, respectively.  
The analogous $w^A_N$ and $w^A_D$ fits for the $ud$ channel are 
shown in Figs. 4 and 5 of Ref. \cite{MKlett01}, respectively. 
The ratio $R= 25.6\pm 2.6$ implied by the above
results is in good agreement with the value, $24.4\pm 1.5$ 
obtained from ChPT in Ref.~\cite{leutwylerqmasses}.

If one repeats the pFESR analysis, but now with the ILM contributions
set to zero, one finds, for the $ud$ case,
\begin{eqnarray}
&&[m_u+m_d](2\ {\rm GeV})\, =\, 9.9\pm 1.2_\Gamma \pm
1.0_{theory}\pm 0.5_{method}\ {\rm MeV}\label{udnoilmmass} \\
&&f_{\pi (1300)}\, =\, 2.41\pm 0.50_{\Gamma}\pm 0.21_{theory}\pm 0.27_{method}
\ {\rm MeV}\label{udnoilmf1} \\
&&f_{\pi (1800)}\, =\,  1.36\pm 0.16_{\Gamma}\pm 0.09_{theory}
\pm 0.11_{method}\ {\rm MeV}\ ,
\label{udnoilmf2}
\end{eqnarray}
and, for the $us$ case,
\begin{eqnarray}
&&m_s(2\ {\rm GeV})\, =\, 116\pm 7_{theory}\pm 3_{method}\ 
{\rm MeV}\label{usnoilmmass} \\
&&f_{K(1460)}\, =\, 22.9\pm 2.1_{theory}\pm 1.2_{method}
\ {\rm MeV}\label{usnoilmf1} \\
&&f_{K(1830)}\, =\, 14.5\pm 1.5_{theory}\pm 0.4_{method}\ {\rm MeV}\ .
\label{usnoilmf2}
\end{eqnarray}
The corresponding OPE/spectral integral match is again excellent.
This is illustrated for the $us$ case, for the $w^A_N$ family of pFESR's, in 
Fig.~\ref{figusNnoILM}. 
(The agreement for the corresponding $w^A_D$
pFESR's as well as that for the $ud$ case is not shown explicitly, but is,
in fact, of equal quality to that for the $us$ $w^A_N$ family.)  
The resulting mass ratio, 
$R=23.3\pm 2.8$, is also in good agreement with that
obtained from ChPT. We thus see that, while 
the pFESR fit provides a good determination of $m_i+m_j$
{\it and} the resonance decay constants {\it once the form of the
theoretical side of the sum rule (i.e., whether including or excluding
ILM contributions) has been fixed}, it does  
not, by itself, provide any additional evidence as to whether 
inclusion or exclusion of these contributions is favored.
While inclusion of ILM effects is, of course,
indicated by arguments external to the pFESR analysis, the pFESR
analysis itself shows only that, in the absence of these contributions,
significantly larger values of the relevant
quark mass combination and second resonance decay constant
are required in both the $ud$ and $us$ channels.

We now turn to the BSR analyses of the $ud$ and $us$ channels,
which should provide additional constraints on the ILM modelling 
of instanton effects in the pFESR analyses.  Expressions
for the Borel transforms of the OPE side of the sum rules can
be found in Refs.~\cite{jm,dps,cps}, and that for the Borel transform
of the ILM contributions in Ref.~\cite{elias97}.
We take central values for all OPE input, and employ the corresponding
central values for the excited resonance decay constants, determined
above, as input to the BSR analysis.
To facilitate the BSR/pFESR comparison, we quote only those
errors present in the BSR analysis which do {\it not} also enter the pFESR
analysis, namely those associated with (1) the $\pm 0.5\ {\rm GeV^2}$
variation of the continuum threshold parameter $s_0$ 
about its optimal stability value and (2) the assumed $20\%$
uncertainty in the size of the continuum spectral contribution.
(Additional errors, associated with uncertainties in
the values of the OPE input parameters, are
common to both analyses, and the corresponding errors, as
a result, are strongly correlated between the pFESR and BSR treatments.)
To be conservative, we take, as our estimates for these 
errors, the maximum change in the value extracted for $m_i+m_j$ 
in our BSR analysis window (see below) produced by the stated variations 
in $s_0$ and the magnitude of the continuum contribution.
These two sources of error have been combined in quadrature in quoting 
results below. 
A conventional rule-of-thumb is that the BSR analysis window
should be restricted to $M^2$ values for which the 
perturbative continuum contribution is
less than $\sim 50\%$ of the OPE contribution (for a discussion see,
for example, Ref.~\cite{leinweber97}).  Since,
for the $ud$ case, this corresponds to $M^2$ less than $\sim 3\ {\rm GeV}^2$,
we work with a BSR analysis window $2\ {\rm GeV}^2\leq M^2\leq 3\ {\rm GeV^2}$.

The dependence of $[m_i+m_j](2\ {\rm GeV})$ on $M^2$
in the extended range $2\ {\rm GeV}^2\leq M^2\leq 4\ {\rm GeV}^2$ resulting
from the BSR ILM analyses is shown in Fig. \ref{usborel} for $ij=us$. 
(The analogous result for $ij=ud$ is shown in Fig. 6 of Ref. \cite{MKlett01}.)
The solid line corresponds to the 
optimal stability value of $s_0$, the upper and lower
lines to values $0.5\ {\rm GeV}^2$ lower and
higher, respectively.
The quark mass values obtained from this analysis are
\begin{eqnarray}
&&[m_u+m_d](2\ {\rm GeV})\, =\, 7.5\pm 0.9\ {\rm MeV}
\label{bsrmumd} \\
&&m_s(2\ {\rm GeV})\, =\, 91 \pm 9\ {\rm MeV}\ .
\label{bsrms} 
\end{eqnarray}
These results are to be compared to the central pFESR values
of Eqs.~(\ref{udilmmass}) and (\ref{usilmmass}) 
above.  The consistency of the two determinations is excellent
for the $ud$ channel, but only marginally acceptable for the $us$
channel.  The consistency of the central
pFESR and BSR $us$ determinations can be improved by
allowing somewhat larger values of $\eta_{us}$.
For example,
$\eta_{us}=0.8$ produces a central value 
$m_s(2\ {\rm GeV})=97\ {\rm MeV}$, with a corresponding
central BSR determination $89\pm 9\ {\rm MeV}$, while
$\eta_{us}=1$ corresponds to $m_s(2\ {\rm GeV})=92\ {\rm MeV}$
(pFESR) and $87\pm 10\ {\rm MeV}$.  In view of the size of the 
BSR errors, such improvement cannot be taken as physically meaningful;
this exercise does, however, indicate that errors comparable in size to
the difference of the pFESR and BSR central values, associated
with the crudeness of the ILM representation of instanton effects,
may still be present in the pFESR results.  
We will therefore include an additional error, given
by this difference of central values, in our final version of
the errors for the light quark masses.

For the case that no ILM contributions are included,
the BSR results, corresponding to central values
of all input, and the corresponding central values of the
resonance decay constants, are
\begin{eqnarray}
&&[m_u+m_d](2\ {\rm GeV})\, =\, 8.8\pm 0.6\ {\rm MeV} \nonumber \\
&&m_s(2\ {\rm GeV})\, =\, 100\pm 6\ {\rm MeV}\ ,
\end{eqnarray}
which are to be compared to the central values
of Eqs.~(\ref{udnoilmmass}) and (\ref{usnoilmmass}).  
The pFESR determinations in both cases lie significantly outside 
the range allowed by the BSR error.

Consistency between pFESR and BSR analyses thus favors
inclusion of the ILM contributions.  
To see that the level of inconsistency
between the pFESR and BSR results in the absence
of ILM contributions is, in fact, significant,
the following exercise is useful.  Rather than
optimizing the pFESR analysis by varying simultaneously $m_i+m_j$,
$f_1$ and $f_2$, we may, for each value of $m_i+m_j$, find
the values of $f_1$ and $f_2$ which produce the best OPE/spectral integral
match.  We then use these values of $f_1$ and $f_2$, as usual,
as input to the corresponding BSR analysis and look for
those values of $m_i+m_j$ for which the pFESR input value
is compatible with the BSR output value, 
within the additional errors of the BSR analysis.

For the $ud$ case, in the absence of ILM contributions,
this compatibility is obtained only for
$[m_u+m_d](2\ {\rm GeV})$ less than $8.1\ {\rm MeV}$
(pFESR value)/$7.6\ {\rm MeV}$ (BSR value).
Taking the ``marginal'' case, corresponding to the 
pFESR value $[m_u+m_d](2\ {\rm GeV})=8.1\ {\rm MeV}$
to be specific, one finds that,
although the quality of the OPE+ILM/spectral
integral match is significantly worse than that for the
fully optimized fit above, it is perhaps still acceptable 
(see Fig.~\ref{udnoILMcompatibility} for the fit quality
for the $w_D$ case; the quality is comparable, though
marginally better, for the $w_N$ case).  Thus, in this case,
although the inclusion of ILM contributions is favored,
we do not consider it possible to rule out their absence.
Note, however, that the analysis, in the absence of ILM
contributions, is only self-consistent for values of
$m_u+m_d$ compatible with those obtained from the
analysis including ILM contributions.  The value of $f_{\pi (1300)}$
obtained in this case, $1.74\ {\rm MeV}$, also turns out
to be compatible, within errors, with that given by Eq.~(\ref{udilmf1}).

For the $us$ case, in the absence of ILM contributions,
compatibility is achieved only for $m_s(2 \ {\rm GeV})$
less than $94\ {\rm MeV}$ (pFESR value)/$89\ {\rm MeV}$ (BSR value).  
The ``best'' fit pFESR solution for such a value of $m_s$, however, 
represents an extremely poor quality OPE+ILM/spectral
integral match{\begin{footnote}{The {\it average} OPE/spectral integral
discrepancy over the $s_0,A$ analysis window is, for example,
$23\%$ for the $w_D$ pFESR family.}\end{footnote}}.
We thus find no acceptable, consistent spectral solution
in the $us$ case without the inclusion of ILM contributions.
This, of course, also favors the inclusion
of such contributions for the $ud$ channel.

In view of these observations, we take as our final results 
those corresponding to the pFESR analysis with direct instanton contributions
estimated using the ILM.  Including our additional estimate of
the error associated with the crudeness of the ILM,
and combining all sources of error in quadrature,
our final results for the light quark masses become
\begin{eqnarray}
&&[m_u+m_d](2\ {\rm GeV})\, =\, 7.8\pm 1.1\ {\rm MeV}\label{mumdff} \\
&&m_s(2\ {\rm GeV})\ =\, 100\pm 12\ {\rm MeV}\ .
\label{finalpFESR}\end{eqnarray}
Since the pFESR/BSR consistency is excellent for the $ud$
channel, but marginal for the $us$ channel, an alternate
determination of $m_s$, using the result of Eq.~(\ref{mumdff}) above
in combination with the ChPT-determined mass ratio $R=24.4\pm 1.5$,
might be preferable.  The result of this determination,
\begin{equation}
m_s(2\ {\rm GeV})\ =\, 95\pm 15\ {\rm MeV}\ ,
\end{equation}
is in good agreement with that of Eq.~(\ref{finalpFESR}), with
only slightly larger errors.
Recall that {\it self-consistent} versions of the combined pFESR/BSR
analysis in which direct instanton contributions are neglected, in fact,
yield values for the light quark masses completely compatible 
with those of Eqs.~(\ref{mumdff}) and (\ref{finalpFESR}).
For the resonance decay constants, we note that, although the value of
the second resonance decay constant is sensitive to whether or not
one includes ILM contributions, that of the first resonance 
is largely insensitive to the presence or absence of ILM
contributions, the central values differing
by considerably less than the uncertainties on the individual
determinations.  We thus believe that, although the ILM may represent a
relatively crude model for implementing direct instanton
effects, the determination of the $\pi (1300)$
and $K(1460)$ decay constants given by Eqs.~(\ref{udilmf1}) and
(\ref{usilmf1}) should be reliable to within the stated errors.
Combining these errors in quadrature we then have, for our final results,
\begin{eqnarray}
f_{\pi (1300)}\, &=&\, 2.20\pm 0.46\ {\rm MeV}\label{udfinalf1} \\
{\rm and} \qquad
f_{K(1460)}\, &=&\, 21.4\pm 2.8\ {\rm MeV}\ .\label{usfinalf1}
\end{eqnarray}
That these values differ by a factor of $\sim 10$ is compatible
with the fact that the excited pseudoscalar decay constants
vanish in the chiral limit, and hence are proportional
to the relevant quark mass combination near that limit.

\subsection{Discussion}

Other recent sum rule analyses exist for the pseudoscalar
$ud$~\cite{bpr,p98} and $us$~\cite{jm,dps} channels.  In addition,
sum rule analyses of the $us$ scalar correlator~\cite{jm,cfnp,cps,j98,kmss},
and of flavor breaking in hadronic $\tau$ 
decay~\cite{km00,ckp98,ALEPHstrange,pp99,kkp002,dhppc00}, have been used 
to extract $m_s$.
In this section
we discuss the relation of our work to that of these earlier references.

For the $ud$ pseudoscalar channel, Ref.~\cite{p98} (P98) represents an update
of Ref.~\cite{bpr}.  (The latter employed 3-loop versions for the
OPE $D=0$ contribution, the running mass and the running coupling, 
P98 the 4-loop versions).  We therefore restrict our
discussion to the latter analysis.  The resonance part of the 
P98 spectral function is of the TRDA form, but rescaled by an
overall factor $1.5$.  Two points should be borne in mind 
regarding the value quoted for $m_u+m_d$ in P98.  The first 
is that the analysis 
is based on FESR's involving the weights $w(s)=1$ and $s$.  For these
weights, however, the corresponding vector isovector channel 
FESR's are {\it not} well-satisfied at the scales employed in 
P98.  (The OPE side has a significantly weaker 
$s_0$ dependence than the spectral integral side, the
latter being obtained, in this case, from
experimental $\tau$ decay data~\cite{kmfesr}.)
The second point is that the ratio of quoted
values for the running mass at scales $1$ and $2$ GeV, 
$m(1\ {\rm GeV})/m(2\ {\rm GeV})=1.31$~\cite{p98},
differs from that, $1.38$, obtained using 4-loop 
running with the central ALEPH determination of $\alpha_s(m_\tau )$
as input.  The results of P98 thus correspond to a smaller value,
$\alpha_s(m_\tau )=0.307$, the effect of which would
be to produce a larger value of
$m_u+m_d$.  One would thus expect a poor match between the
OPE and spectral integral sides of pFESR's employing the P98
spectral ansatz and central $m_u+m_d$ value in combination with current
central values for the OPE input.  This is confirmed by
the results shown in Figs.~\ref{bprNps} and \ref{bprDps}, 
which correspond, respectively, to the output from the $w^A_N$ and $w^A_D$ 
pFESR weight family analyses, in our $s_0,A$
analysis window, obtained using the P98 central value for
$m_u+m_d$ and the P98 spectral ansatz.
If one performs a re-analysis, still using the P98 spectral
ansatz, but now optimizing the
value of $m_u+m_d$ using the pFESR approach,
one finds, using central values for all OPE input, 
and including ILM contributions,
\begin{equation}
[m_u+m_d](2\ {\rm GeV})=6.8\ {\rm MeV}\ .
\end{equation}
The same analysis, without ILM contributions, similarly yields
\begin{equation}
[m_u+m_d](2\ {\rm GeV})=7.3\ {\rm MeV}\ .
\end{equation}
Both of these values are, in fact, in agreement with those corresponding
to the upper part of the $s_0$ range displayed in Fig.~2 of P98, though
not with those for $s_0\sim 2\ {\rm GeV}^2$.
In both cases, however, the quality of the OPE(+ILM)/spectral integral match
is much inferior to that obtained obtained using
the solutions for $m_u+m_d$, $f_{\pi (1300)}$ and $f_{\pi (1800)}$
above.  The optimized match is significantly better when
ILM contributions are included than when they are not. However, in spite of 
optimization, the consistency between the $w^A_N$ and $w^A_D$ pFESR families
is not good for the P98 spectral ansatz: as shown in \cite{MKlett01}, the match
for $w^A_N$ is best where that for $w^A_D$ is worst, and vice versa 
(see Figs. 2,3 in \cite{MKlett01}).

For the $us$ pseudoscalar channel, the BSR analyses of Refs.~\cite{jm} 
(JM) and \cite{dps} (DPS) both employ a TRDA construction for
the $K(1460)$ and $K(1830)$ contributions to the spectral ansatz, but
differ in their assumptions about the relative sizes of
the two resonance decay constants: JM assume
$f_2^2 m_2^4/f_1^2 m_1^4 =0.25$, DPS that the spectral
contributions of the two resonances at threshold are
approximately equal (for PDG2000 values of the masses
and widths, this corresponds to 
$f_2^2 m_2^4/f_1^2 m_1^4 \simeq 1.8$).  
The two analyses also differ in their treatment of the theoretical
side, JM employing 3-loop expressions on the OPE side and
DPS the 4-loop expressions which become available subsequent to
the publication of the JM paper.  

We have updated the JM BSR analysis
to include 4-loop contributions to the running mass,
coupling and $D=0$ OPE term.  For OPE input we use the 
values employed in our analyses above.  Including ILM contributions on the 
theoretical side of the BSR, we then find that
the JM spectral ansatz, $\rho_{JM}$, corresponds to 
\begin{equation}
m_s(2\ {\rm GeV})= 96\pm 7\ {\rm MeV}\ .
\label{jmbsrilmused}\end{equation}
Neglecting instanton contributions, as in JM, we obtain instead
\begin{equation}
m_s(2\ {\rm GeV})= 98\pm 6\ {\rm MeV}\ .
\label{jmbsrnoilmused}\end{equation}
(The errors in these equations have the same meaning as those
for the BSR analyses above.)  
If, however, we employ $\rho_{JM}$, as input, not to a BSR
analysis, but to our usual pFESR analysis, 
we find for our central values 
\begin{equation}
m_s(2\ {\rm GeV})= 107\ {\rm MeV}
\end{equation}
if ILM contributions are included, and
\begin{equation}
m_s(2\ {\rm GeV})= 111\ {\rm MeV}
\end{equation}
if they are not.  The fit quality for the optimized match is
rather poor when ILM contributions
are included, but is quite good when they are not.  The latter
point is illustrated
for the $w^A_N$ family of pFESR's in Fig.~\ref{spsjmN} 
(the quality of the match for the $w^A_D$ family, which is not shown, 
is even better).  Despite the existence of both a
good quality pFESR OPE/spectral integral match and an excellent BSR stability 
window, however, we see that the no-ILM pFESR and BSR $m_s$ 
determinations based on $\rho_{JM}$ are inconsistent, 
just as was the case for the determinations associated with the
spectral ansatz based on the values of
$f_{K(1460)}$ and $f_{K(1830)}$ obtained from the no-ILM pFESR analysis.
This is, in fact, not surprising, since the optimized pFESR
spectral ansatz turns out to be rather similar to $\rho_{JM}$,
the $K(1460)$ decay constants of the two models, for
example, differing by less than $6\%$.  

In discussing the DPS analysis of the $us$ pseudoscalar channel,
one should bear in mind that
the result quoted by DPS, $m_s(1\ {\rm GeV})=155\pm 25\ {\rm MeV}$
corresponds to (1) an average of the values obtained
using $\Lambda_{QCD}^{(3)}=280\ {\rm MeV}$ and $380\ {\rm MeV}$, 
(2) an average over values associated with
a range of $s_0$, and (3) neglect of $m_u$ in the 
OPE prefactor{\begin{footnote}{Restoring 
$m_u$ to the prefactor, using ChPT values for the
quark mass ratios, and converting to the scale $\mu =2$ GeV, the
DPS result becomes $m_s(2\ {\rm GeV})=109\pm 18\ {\rm MeV}$.}\end{footnote}}.
Since the choice $\Lambda_{QCD}^{(3)}=280\ {\rm MeV}$ is not
consistent with the ALEPH determination of $\alpha_s(m_\tau)$,
we restrict our attention to the DPS results obtained using 
$\Lambda^{(3)}_{QCD}=380\ {\rm MeV}$, which corresponds very
closely to the central ALEPH determination.
Restoring $m_u$ in the overall OPE prefactor,
and reading off from Fig.~2 of DPS, concentrating on the curve
corresponding to $s_0=6\ {\rm GeV}^2$, which displays the
best stability of $m_s$ with respect to $M^2$, 
the central DPS BSR determination becomes
$m_s(2\ {\rm GeV})= 97\ {\rm MeV}$.  Since the 
details of the spectral ansatz employed are
not fully specified in DPS, we are unable to quote errors equivalent to 
those of our BSR analyses above.  If, however,
we fix the ratio of decay constants in such a
way as to ensure exact equality
of the $K(1460)$ and $K(1830)$ contributions 
to the spectral function at physical threshold, and
neglect ILM contributions, as in DPS, we find that, after
performing our usual pFESR analysis,
the resulting spectral ansatz, run through a BSR analysis
with $s_0=6\ {\rm GeV}^2$, reproduces the DPS central
value exactly.  Estimating our BSR errors as for the
analyses above, we then have, for our DPS-like BSR determination, 
\begin{equation}
m_s(2\ {\rm GeV})= 97 \pm 6\ {\rm MeV}\ .
\end{equation}
The pFESR OPE/spectral integral match corresponding to
this BSR determination is reasonable (see Fig.~\ref{dpsnoilmN} 
for the $w^A_N$ family case; the fit quality for the 
$w^A_D$ family is not shown, but is better than for the
$w^A_N$ family).
The central no-ILM pFESR $m_s$ value,
\begin{equation}
m_s(2\ {\rm GeV})= 109\ {\rm MeV}\ ,
\end{equation}
however, is again inconsistent with the corresponding
BSR value.  The situation is not improved by including ILM
contributions:  re-doing the pFESR analysis, still with the constrained form 
of the spectral ansatz, but now incorporating ILM contributions
on the theoretical side, one finds a poor quality
optimized OPE+ILM/spectral integral match.

We conclude this section with a reminder of the
values obtained for $m_s$ via sum rule analyses of other channels.
Recent treatments of the correlator of 
$\partial_\mu \left(\bar{s}\gamma^\mu u\right)$~\cite{cfnp,j98,kmss},
for which the low-$s$ part of the spectral function is constrained
by $K\pi$ phases, yield values of $m_s(2\ {\rm GeV})$ in the range
$115\pm 25\ {\rm MeV}$, compatible with either the ILM or no-ILM
results above.  Assumptions about the form of the Omnes representation
of the timelike scalar $K\pi$ form factor, and the behavior of the
$K\pi$ phase in the region $s>2.9\ {\rm GeV}^2$, where experimental
phase data does not exist, however,
enter the construction of the spectral function used in those
analyses, so that a significant theoretical systematic error 
is present, in addition to the errors quoted in Refs.~\cite{cfnp,j98,kmss}.
A much cleaner approach, in principle, is the extraction of $m_s$
via pFESR analyses of the flavor-breaking difference of $ud$ and
$us$ vector-plus-axial-vector correlator sums.  
The hadronic spectral function required in this case is measurable in
hadronic $\tau$ decay.  There are two basic complications, 
first, that the OPE representation
of the longitudinal contribution to the $\tau$ hadronic decay width is very
badly behaved at those scales which are kinematically
allowed~\cite{kmtauprob,pp98,kkp001,kmlongprob01} and, second,
that, because of the rather strong cancellation in the $ud$-$us$ spectral
difference, the extracted value of $m_s$ is quite sensitive
to even $\sim 1\%$ uncertainties in the value
of $\vert V_{us}\vert$.  The first problem can be handled by
appropriate weight choices~\cite{km00}.  The second is numerically
relevant because the central
values of the determinations of $\vert V_{us}\vert$
based on (1) experimental $K_{\ell 3}$ data, 
$\vert V_{us}\vert = 0.2196\pm 0.0023$ 
and (2) CKM unitarity, in combination with the experimental
value of $\vert V_{ud}\vert$, 
$\vert V_{us}\vert =0.2225\pm 0.0035${\begin{footnote}{The central 
value and ``errors'' 
quoted here correspond to the mid-point and extent of the PDG2000
unitarity-constrained fit range.}\end{footnote}}, 
while consistent within errors, differ by $\sim 1.3\%$.  
There has also been some confusion in the literature resulting from
the use, in the various recent theoretical analyses, 
of three different sets of values for the weighted $ud$-$us$ 
spectral differences, corresponding to three different
values of $B_{us}$, the total ($V+A$) branching
fraction into strange hadronic 
states{\begin{footnote}{These three values, which
are in the ratios $1:1.04:1.05$ correspond to (1) the preliminary
(1998) ALEPH analysis of strange decay modes~\cite{cdh98},
(2) the final (1999) version of this analysis~\cite{ALEPHstrange},
and (3) the recent update (2000) reported by Davier~\cite{dhppc00}.
Larger values of $B_{us}$ correspond to smaller values of
the $ud$-$us$ difference, and hence to lower values of $m_s$.}
\end{footnote}}.  The strong $ud$-$us$ cancellation 
makes the extracted value of $m_s$ quite sensitive to 
the (apparently rather small) differences between these $B_{us}$
values.  The discrepancies between the various values of $m_s$ reported 
in the literature, all of which are nominally based on 
the ``same'' (ALEPH) $\tau$ decay data, 
turn out to be almost entirely a reflection of this sensitivity.
The situation is discussed in some detail in Ref.~\cite{rgkm00}, where
the various analyses have also been updated to reflect
the current experimental situation (as reported 
in Ref.~\cite{dhppc00}). Once common input is employed, all hadronic $\tau$
determinations of $m_s$ are in excellent agreement~\cite{rgkm00}.  
The dominant uncertainty remains that associated with $\vert V_{us}\vert$.
Using central values of $\vert V_{ud}\vert$ and $\vert V_{us}\vert$
corresponding to either 
(1) the PDG2000 best independent individual determinations (CKMN)
($\vert V_{ud}\vert =0.9735$ and $\vert V_{us}\vert =0.2196$) 
or (2) the PDG2000 unitarity-constrained fit (CKMU) 
($\vert V_{ud}\vert =0.9749$ and $\vert V_{us}\vert =0.2225$),
one obtains
\begin{equation}
m_s(2\ {\rm GeV})=101\pm 18\ {\rm MeV}\ \ {\rm (CKMN)}\ ,
\end{equation}
and
\begin{equation}
m_s(2\ {\rm GeV})=114\pm 16\ {\rm MeV}\ \ {\rm (CKMU)}\ 
\end{equation}
respectively~\cite{rgkm00}.  Either of these results is compatible
with that obtained from the pseudoscalar channel
analyses above.  

\section{Conclusions}

We have determined $m_u+m_d$, $m_s$, and the decay constants
of the $\pi (1300)$ and $K(1460)$ with good accuracy
from a combined pFESR/BSR study of 
the $ud$ and $us$ pseudoscalar correlators.
Our results show that it is important to require the
consistency of the two different sum rule approaches.
Indeed, we have seen that there exist ans\"atze for
the hadronic spectral functions which produce
both extremely good BSR stability plateaus and high-quality
pFESR OPE(+ILM)/spectral integral matches, but for
which the output quark mass combinations are inconsistent.
This means that BSR or pFESR treatments, by themselves,
do not provide sufficiently strong constraints to allow one to
simultaneously constrain the unknown quark masses, unknown resonance decay 
constants and the theoretical modelling of direct instanton effects.
The combination of the two approaches does, however,
provide sufficiently strong constraints.  The consistency of
the combined analysis is particularly compelling for the
$ud$ case.  The values obtained for the light quark masses
are in excellent agreement with determinations from other
sources, giving us further confidence in the reliability
of the combined analysis.  The corresponding determinations
of the $\pi (1300)$ and $K(1460)$ decay constants
are accurate to $20\%$ and $10\%$ respectively.
The latter determination is relevant to future improvements
in the extraction of $m_s$ from hadronic $\tau$ decay data.
While B factory data will
dramatically reduce the errors on the experimental $us$
vector-plus-axial-vector $\tau$ decay distribution, 
the ability to use this improvement to
significantly reduce the errors on the corresponding determination
of $m_s$ will depend on one's ability to work with pFESR's
involving weights for which the $ud$-$us$ cancellation is
significantly reduced, so that the errors resulting
from the uncertainty in $\vert V_{us}\vert$ will,
as a result, play a significantly reduced role.  
The existence of significant theoretical systematic uncertainties
in versions of this analysis which include longitudinal OPE 
contributions~\cite{kmlongprob01} means that 
``non-inclusive'' analyses (involving only the sum of
spin $0$ and $1$ correlator components) will eventually be required.
A knowledge of the decay constants of the excited strange
pseudoscalar and scalar resonances allows a straightforward
subtraction of the longitudinal contributions to the experimental
distributions.  In the absence of an experimental spin separation,
sum rule determinations of the strange scalar and
pseudoscalar resonance decay constants with an accuracy even
a factor of three worse than that obtained above for the
$K(1460)$ are already extremely useful as input for such non-inclusive
analyses.

\acknowledgements
KM would like to acknowledge the ongoing support of the Natural Sciences and
Engineering Research Council of Canada, and also the hospitality of the
Theory Group and TRIUMF and the Special Research Centre for the 
Subatomic Structure of Matter at the University of Adelaide.  
JK would like to acknowledge partial support from Schweizerischer 
Nationalfonds and EC-Contract No. ERBFMRX-CT980169 (EURODA$\Phi$NE). 
\vfill\eject

\begin{figure}
\centering{\
\psfig{figure=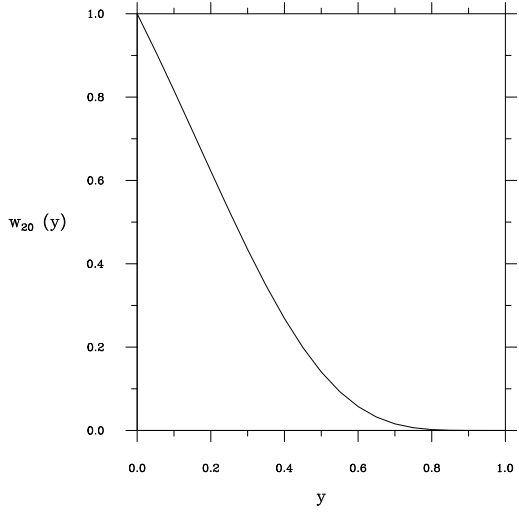,height=7.5cm}}
\vskip 0.15in
\caption{The behavior of the weight $w_{20}(y)$ in the 
pFESR integration region.}
\label{diagrams}
\end{figure}



\vskip .5in
\begin{figure}
\centering{\
\psfig{figure=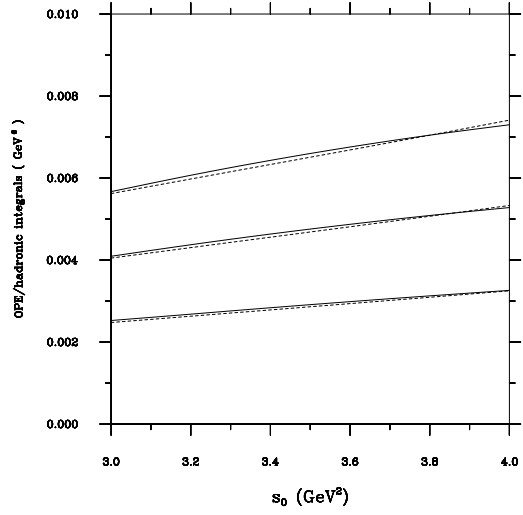,height=7.5cm}}
\vskip 0.15in
\caption{The OPE+ILM versus 
hadronic (spectral integral) sides of the $us$ $w_N^A$ family
of pFESR's, for $m_s+m_u$, $f_{K(1460)}$ and $f_{K(1830)}$
given by the central values of Eqs.~(\ref{usilmmass}),
(\ref{usilmf1}) and (\ref{usilmf2}).  
The solid lines are
the hadronic integrals, the dashed lines the corresponding OPE
integrals.  The lower, middle and upper lines in each case
correspond to $A=0,2$ and $4$, respectively.}
\label{figusNILM}
\end{figure}

\vskip .5in
\begin{figure}
\centering{\
\psfig{figure=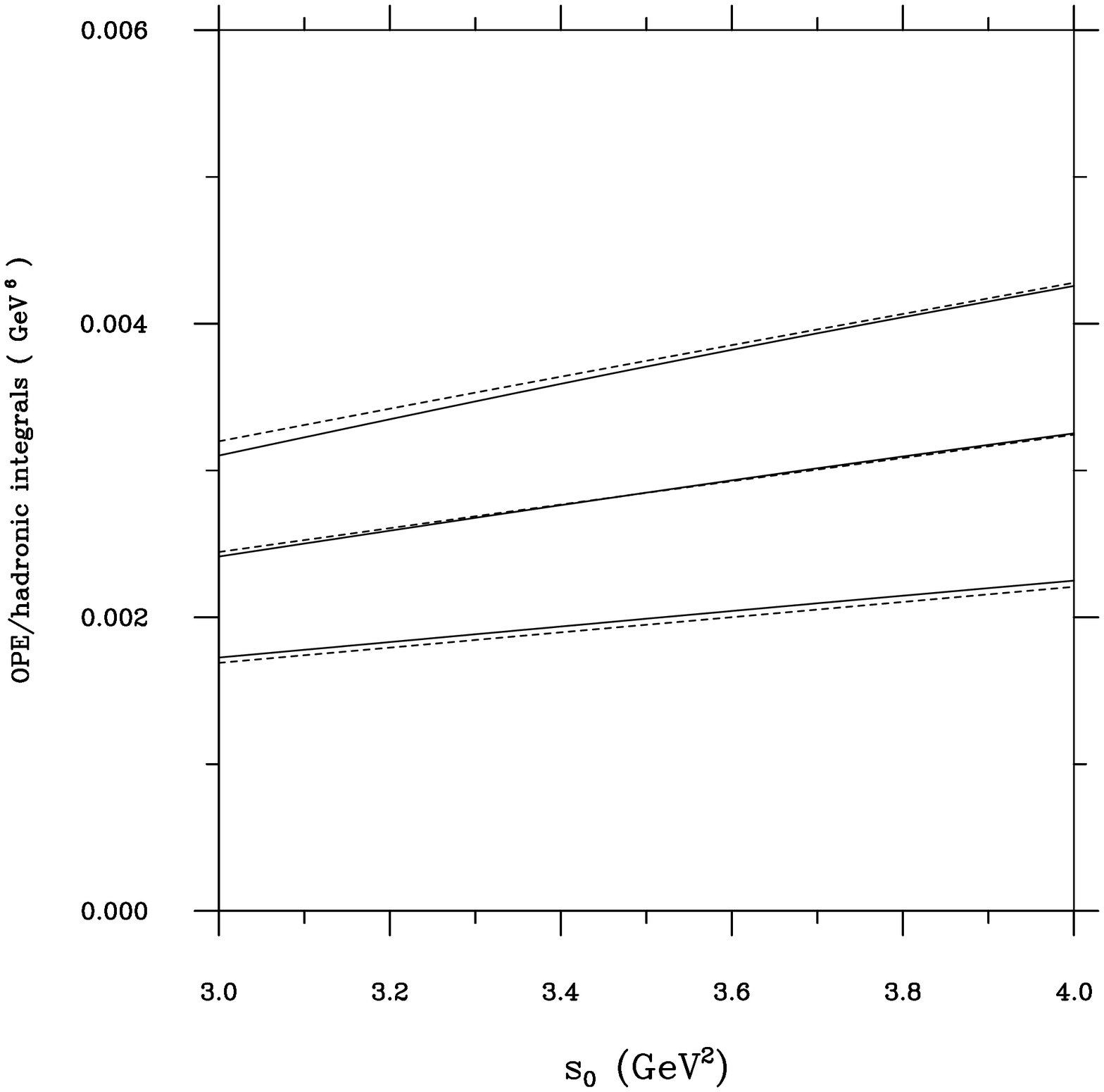,height=7.5cm}}
\vskip 0.15in
\caption{The OPE+ILM versus 
hadronic (spectral integral) sides of the $us$ $w_D^A$ family
of pFESR's for $m_s+m_u$, $f_{K(1460)}$ and $f_{K(1830)}$
given by the central values of Eqs.~(\ref{usilmmass}),
(\ref{usilmf1}) and (\ref{usilmf2}).  
The identification of
OPE and hadronic integrals, and the cases $A=0,2,4$ is
as for Fig.~\ref{figusNILM} above.}
\label{figusDILM}
\end{figure}


\vskip .5in
\begin{figure}
\centering{\
\psfig{figure=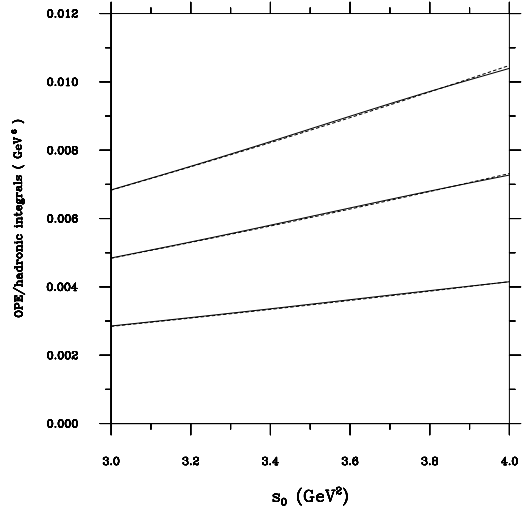,height=7.5cm}}
\vskip 0.15in
\caption{The OPE versus 
hadronic (spectral integral) sides of the $us$ $w_N^A$ family
of pFESR's for $m_s+m_u$, $f_{K(1460)}$ and $f_{K(1830)}$
given by the central values of Eqs.~(\ref{usnoilmmass}),
(\ref{usnoilmf1}) and (\ref{usnoilmf2}), i.e., in the
absence of ILM contributions.  
The identification of
OPE and hadronic integrals, and the cases $A=0,2,4$ is
as for Fig.~\ref{figusNILM} above.}
\label{figusNnoILM}
\end{figure}
\vskip .5in


\vskip .5in
\begin{figure}
\centering{\
\psfig{figure=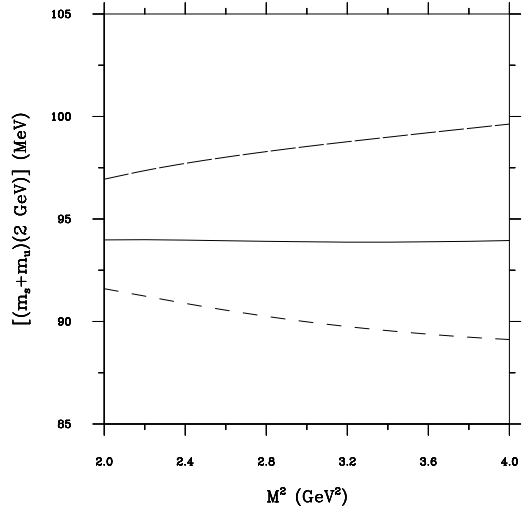,height=7.5cm}}
\vskip 0.15in
\caption{The value of $[m_s+m_u](2\ {\rm GeV})$, as a function
of the square of the Borel mass, $M^2$, 
extracted from the BSR analysis of the $us$
pseudoscalar correlator described in the text.  The solid
line corresponds to $s_0=4.22\ {\rm GeV}^2$, which produces
optimal stability for $m_s+m_u$ with respect to $M^2$ in the window 
$2\ {\rm GeV}^2\leq M^2\leq 3\ {\rm GeV}^2$.
The lower (short) dashed line corresponds
to $s_0=4.72\ {\rm GeV}^2$ and the upper
(long) dashed line to $s_0=3.72\ {\rm GeV}^2$.}
\label{usborel}
\end{figure}

\vskip .5in
\begin{figure}
\centering{\
\psfig{figure=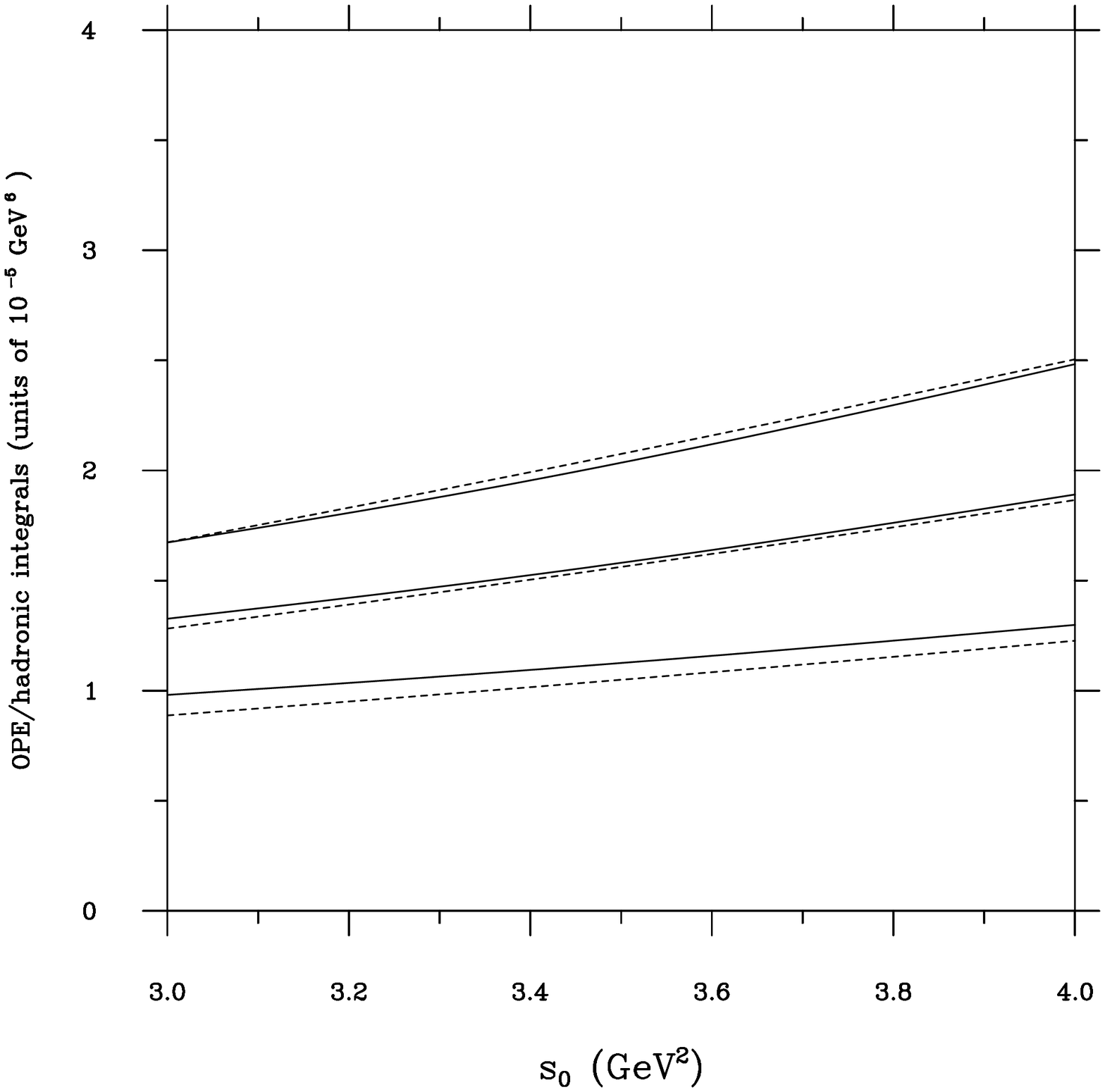,height=7.5cm}}
\vskip 0.15in
\caption{The $ud$ OPE/spectral integral match obtained
for the $w_D^A$ pFESR family using as pFESR input the value
$[m_u+m_d](2\ {\rm GeV})=8.1\ {\rm MeV}$, the {\it largest}
pFESR input for which pFESR and 
BSR values of $m_u+m_d$ are consistent.
All notation is as for the pFESR figures above.  This largest
``marginal'' $m_u+m_d$ value produces the best OPE/spectral integral
match among those input values for which the pFESR input and BSR
output values are consistent; the fit quality, moreover, deteriorates 
rapidly as one goes to lower values of the pFESR input.}
\label{udnoILMcompatibility}
\end{figure}

\vskip .5in
\begin{figure}
\centering{\
\psfig{figure=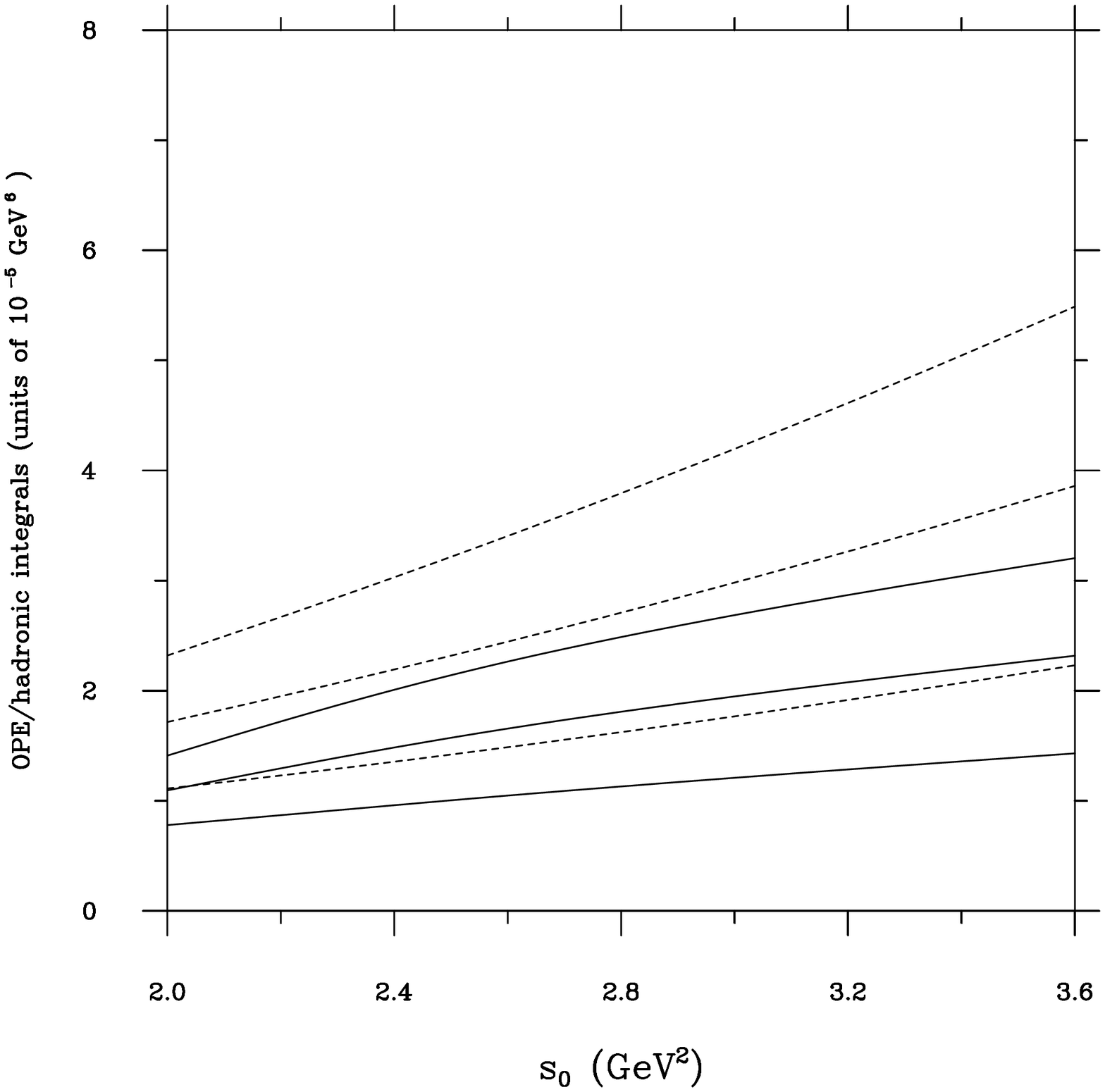,height=7.5cm}}
\vskip 0.15in
\caption{The $ud$ OPE/spectral integral match obtained
for the $w_N^A$ pFESR family using the
central values of all OPE input, the quoted P98 value of 
$[m_u+m_d](1\ {\rm GeV})$ and the P98 spectral ansatz.
All notation is as for the pFESR figures above.}
\label{bprNps}
\end{figure}

\vskip .5in
\begin{figure}
\centering{\
\psfig{figure=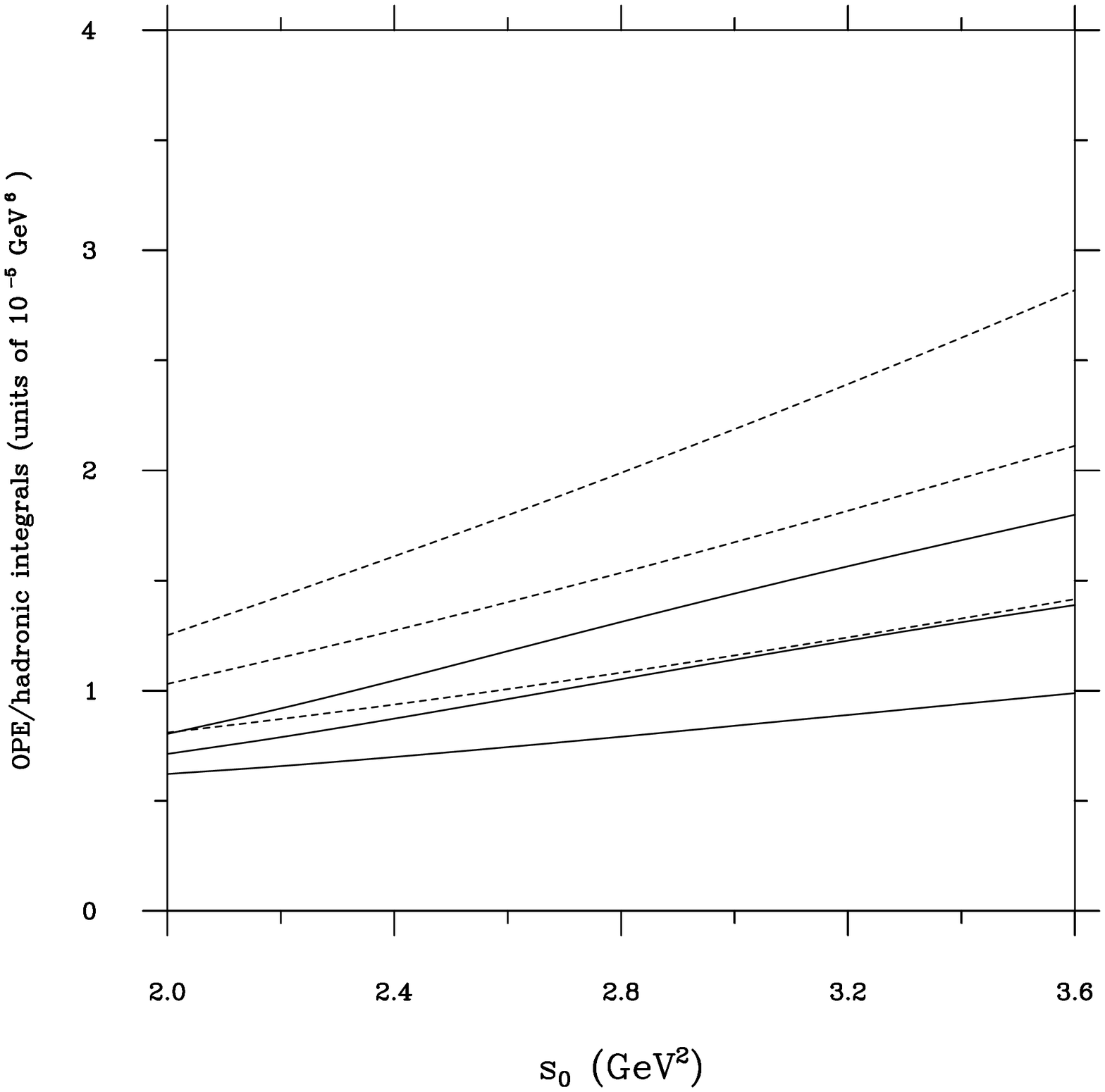,height=7.5cm}}
\vskip 0.15in
\caption{The $ud$ OPE/spectral integral match obtained
for the $w_D^A$ pFESR family using the
central values of all OPE input, the quoted P98 value of 
$[m_u+m_d](1\ {\rm GeV})$ and the P98 spectral ansatz.
All notation is as for the pFESR figures above.}
\label{bprDps}
\end{figure}



\vskip .5in
\begin{figure}
\centering{\
\psfig{figure=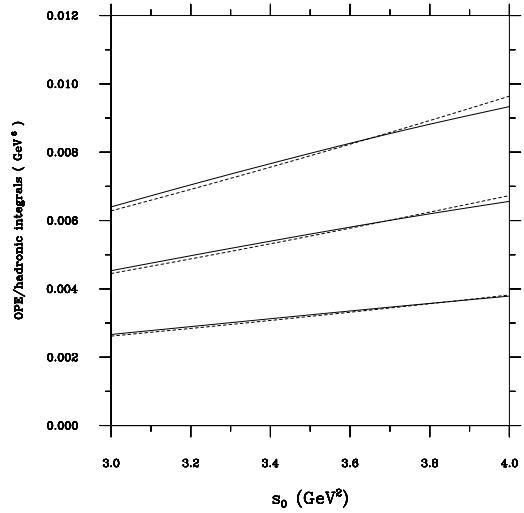,height=7.5cm}}
\vskip 0.15in
\caption{The $us$ pseudoscalar OPE/spectral integral match obtained
for the $w_N^A$ pFESR family using the JM spectral ansatz,
central values of all OPE input, and no ILM contributions,
after optimization of $m_s+m_u$ in a combined $w_N^A$, $w_D^A$
pFESR analysis.  All notation is as for the pFESR figures above.}
\label{spsjmN}
\end{figure}

\vskip .5in
\begin{figure}
\centering{\
\psfig{figure=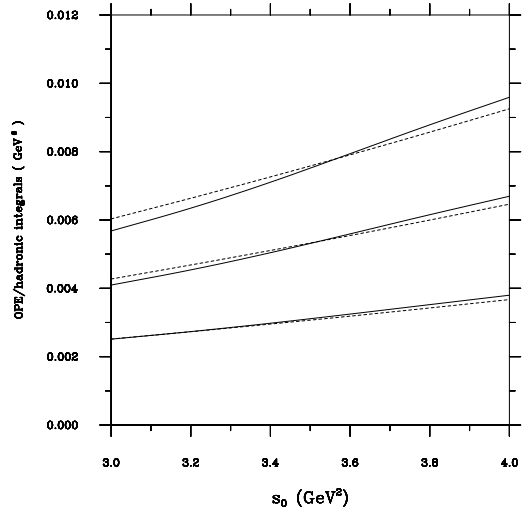,height=7.5cm}}
\vskip 0.15in
\caption{The $us$ pseudoscalar OPE/spectral integral match 
for the $w_N^A$ pFESR family involving the spectral ansatz
obtained from a combined $w^A_N$, $w^A_D$ pFESR analysis
after imposing the DPS-like constraint on the ratio of $K(1460)$
and $K(1830)$ decay constants.  The results correspond to
central values of all OPE input, and to neglect of ILM contributions.
All notation is as for the pFESR figures above.}
\label{dpsnoilmN} 
\end{figure}

\vfill\eject

\end{document}